\documentclass[useAMS,usenatbib,fleqn] {mnras}
\usepackage{epsfig}
\usepackage{amsmath}
\usepackage{graphicx}
\usepackage{amssymb} 
\usepackage{pdflscape}
\usepackage{epstopdf}

\title[The origin of 2nd-generation stars in GCs]{The merger of hard binaries in globular clusters as the primary channel for the formation of second generation stars.}

\author[Kravtsov et al.]{Valery Kravtsov$^{1}$\thanks{E-mail, VK: vkravtsov1958@gmail.com}, Sami Dib$^{2}$, Francisco A. Calder\'{o}n$^{3}$\\
$^{1}$Sternberg Astronomical Institute, Lomonosov Moscow State University, University Avenue 13, 119899 Moscow, Russia\\
$^{2}$Max Planck Institute for Astronomy, K\"{o}nigstuhl 17, D-69117, Heidelberg, Germany\\
$^{3}$Departamento de F\'{i}sica, Universidad Cat\'{o}lica del Norte, Av. Angamos 0610, Antofagasta, Chile\\
}

\begin{document}
\maketitle

\date{Accepted XXX. Received XXX}

\pagerange{\pageref{firstpage}--\pageref{lastpage}}
\pubyear{2016}
\label{firstpage}

\begin{abstract}  
We have recently presented observational evidence which suggests that the origin of the second generation (G2) stars in globular clusters (GCs) is due to the binary-mediated collision of primordial (G1) low-mass main-sequence (MS) stars. This mechanism avoids both the mass budget problem and the need of external gas for dilution. Here, we report on another piece of evidence supporting this scenario: (1) the fraction of MS binaries is proportional to the fraction of G1 stars in GCs and, at the same time, (2) the smaller the fraction of G1 stars is, the more deficient binaries of higher mass ratio (q$>0.7$) are. They are, on average, harder than their smaller mass-ratio counterparts due to higher binding energy at a given primary mass. Then (2) implies that (1) is due to the merging\slash collisions of hard binaries rather than to their disruption. These new results complemented by the present-day data on binaries lead to the following conclusions: (i) the mass-ratio distribution of binaries, particularly short-period ones, with low-mass primaries, $M_{\rm P} < 1.5$ M$_{\sun}$, is strongly peaked close to q$=1.0$, whereas (ii) dynamical processes at high stellar density tend to destroy softer binaries and make hard (nearly) twin binaries to become even harder and favor their mergers and collisions. G2 stars formed this way gain mass that virtually doubles the primary one, $2M_{\rm P}$, at which the number of G1 stars is $\sim5$ times smaller than at $M_{\rm P}$ according to the slope of a Milky Way-like IMF at $M_{\rm MS} < 1.0$ M$_{\sun}$.

\end{abstract} 


\section{INTRODUCTION}\label{introduc}

Globular clusters (GCs) with a present-day mass exceeding a threshold value of about $M_{\rm GC} \sim 10^{4}$ M$_{\sun}$ \citep{Carrettaetal2010a,Bragagliaetal2017,Tangetal2021} are known to be typically composed of two sub-populations or generations of stars, a primordial generation (G1) and a secondary one (G2). Whereas the surface abundance patterns of G1 red giant branch (RGB) stars have been spectroscopically established to be similar to those of the field counterparts of the same iron abundance, G2 stars have been found \citep[e.g.,][and references therein]{Carretta2009} to have distinct abundances of light elements (C, N, O, Na, Mg, Al) and key (anti)correlations between them. Namely, the abundances of N, Na, Al are increased, while those of C, O, Mg are decreased. However, there are important nuances. Specifically, \citet{Pancinoetal2017} and \citet{Nataf_etal2019} found that the Mg--Al anti-correlation tend to disappear in less massive or most metal-rich GCs. \citet{Nataf_etal2019} argue that this suggests more than one chemical polluter contributed to the chemical manifestations of the multiplicity in GCs. The presence of sub-populations of G2 stars with such chemical characteristics depending on cluster parameters is specific to GCs with no evident counterparts in open star clusters \citep[see, however,][for possible exceptions]{Pancino2018} nor in the field. An exception is the Galactic bulge \citep{Schiavonetal2017} which is characterized by a high stellar density. The presence of G2 stars in the Galactic halo is not excluded \citep{martellgrebel2010}, in particular, due to their escape from GCs.

Outside of the Milky Way, the manifestations typical of multiple stellar populations in Galactic GCs were revealed in massive star clusters older than 2 Gyr in the Large and Small Magellanic Clouds \citep[e.g.,][]{Hollyheadetal2017,Niederhoferetal2017,Hollyheadetal2018,Milonetal2020,Lietal2021,martocchiaetal2021}. More details on this particular subject and on the obtained results can be found in \citet{Salgadoetal2022}. Moreover, young and intermediate-age massive star clusters in the the Magellanic Clouds exhibit a complexity of their stellar populations in the form of an extended MS turnoff \citep{Mackey_BrobyNiel2007,Mackeyetal2008,Milonetal2009}, which is believed to be another manifestation of the multiplicity at the respective age.

Abundant knowledge about multiple stellar populations in GCs have been obtained thanks to observational and theoretical investigations carried out to date. The key facts and various aspects concerning multiple populations in GCs, including massive star clusters in the Magellanic Clouds, have been reviewed in dedicated papers, for instance, by \citet{BastianLardo2018}, \citet{Grattonetal19}, and \citet{MiloMari2022}. In spite of gaining a deeper insight, the origin of G2 stars paradoxically remains a controversial issue. New findings make the problem even more complex and confusing. Very special assumptions are sometimes required to explain some observations. This could be attributed to the shared idea between virtually all of the proposed scenarios about the nature of the majority of G2 stars, namely the canonical view that G2 stars are formed from gas lost by intermediate- and high-mass G1 stars. 

We suggest that the collision\slash merging of G1 stars might be a plausible alternative for explaining the origin of G2 stars, by relying on observational evidence. \citet{Kravtsovetal2022} have recently examined available data on both the slope of the present-day mass function ($\alpha_{\rm pd}$, MF) and the fraction of G1 RGB stars ($N_{\rm G1}/N_{\rm tot}$) for a sample of 35 Galactic GCs. They found that an anti-correlation between the two quantities is statistically significant for GCs of higher mass ($M_{\rm GC} > 10^{5.3}$ M$_{\sun}$). Also, a very significant anti-correlation has been found between ($N_{\rm G1}/N_{\rm tot}$) and the stellar encounter rate for a sample of 54 GCs. Taken together, these relationships imply that the origin of G2 stars presently observed in GCs [i.e., with their (progenitor) mass $M_{\rm MS} < 0.9$ M$_{\sun}$] could be due to binary-mediated collision\slash merging of MS G1 stars originally belonging to the low-mass end ($[0.1-0.5]$ M$_{\sun}$) of the initial mass function (IMF). We also considered a generic model \citep{Kravtsovetal2022,Dibetal2022} that showed the feasibility of such a mechanism to produce (1) a sufficient fraction of G2 stars and (2) the general trend of the dependence of ($N_{\rm G1}/N_{\rm tot}$) on $\alpha_{\rm pd}$. An increased dispersion of points in the latter dependence can likely be caused by variations in the IMF \citep{Dibetal2022} and a higher mass loss from lower-mass GCs. 

Our approach highlights the potentially important or even decisive contribution to the formation of G2 stars via this collisional mode of star (trans)formation. While none or negligible (trans)formation of stars from the coalescence of other stars occurs in low to intermediate density environments, this mode can play a much more prominent role at high stellar densities such as those that are thought to be prevailing in GCs at their infancy \citep{Dib2023}. Moreover, it has been argued since long ago \citep[e.g.,][and references therein]{Meyl_Hegg1997} that binary mediated stellar encounters are more probable among other possible channels of stellar collision. More recently, by modeling stellar encounters relevant to the conditions in the Galactic nuclear star cluster, \citet{MastrobuChurDav2021} found that collisions between pairs of MS stars are the most common stellar encounters leading to mergers. 

In the context of the nature of multiple stellar populations, a particular attention is given to the established difference of the occurrence of G2 stars in the three above-mentioned types of environments, i.e. GCs, open star clusters, and Galactic field. Interestingly, this is not the only dissimilarity between stellar populations in GCs, on the one hand, and the field and open clusters, on the other. The fraction of the presently observed binaries in GCs is notably smaller than in both the field and open star clusters of similar metallicity \citep[e.g.,][among others]{BicaBonatto2005,Sollimetal2007,milonetal2012a,Raghavanetal2010,Tokovinin2014b,AlbrowUlusele2022,Cordonetal2023}. Oddly enough, this important fact is not normally associated with the origin of G2 stars. We believe that the reason for this is twofold. First, it is naturally believed that the (much) smaller fractions of binary stars in GCs is fully due to their destruction in the high-density environment. Second, the scenarios of the formation of G2 stars from G1 stars ejecta do not make any particular room for the issue of binarity. On the contrary, in the framework of the collision\slash merger-based approach, the smaller fraction of MS binaries in GCs is very relevant, particularly by comparing some details of the present-day data, both theoretical and observational, on binary stars in different environment with the results obtained here. Note, however, that the issue of binaries and their occurrence in different environment is really complicated.

The present paper is organized in the following way. In the next section (Section~\ref{data}), we describe the data used for our analysis. In Section~\ref{results}, we show the existence of a tight correlation between the fraction of binary stars and the fraction of G1 RGB stars in GCs, ($N_{\rm G1}/N_{\rm tot}$). A supporting evidence that this correlation is not due to the disruption of binaries but rather to their merger is also presented. It is twofold: (i) a trend shown between the same parameter ($N_{\rm G1}/N_{\rm tot}$) and the relative fraction of higher mass-ratio binaries, and (ii) the present-day data characterizing them as high-mass ratio hard binaries with fate as mergers. We discuss our findings and their implications in Section~\ref{discussion}, along with our conclusions.

\section{DATA} \label{data}

In the present study, we rely on publicly available sets of data on Galactic GCs and their stellar populations (SPs). One of these is the atlas composed by \citet{Milone_etal2017} on multiple SPs in 57 GCs with previously obtained HST photometry in different passbands \citep{Sarajedini_etal2007,Piotto_etal2015} in the central parts of the GCs of this sample. The authors isolated RGB stars, separated them into subpopulations belonging to different generations, and deduced a set of parameters characterizing each generation. Specifically, \citet{Milone_etal2017} quantified the relative number of G1 red giants ($N_{\rm G1}/N_{\rm tot}$) as the ratio between the number of G1 ($N_{\rm G1}$) and the total number of red giants in the observed area of each GC. \citet{Milone_etal2017} have been able to estimate the ($N_{\rm G1}/N_{\rm tot}$) parameter in 54 of the 57 GCs.

Another set of data is on the fraction of binaries derived by \citet{milonetal2012a} in 59 Galactic GCs. The list of these GCs essentially overlaps with the one for which the parameter ($N_{\rm G1}/N_{\rm tot}$) is measured. However, the final list of GCs in common was defined by the following details concerning the data on binaries. The fraction of binaries were estimated and listed for three projected GC areas confined between radial distances from the cluster centers corresponding to the GC structural parameters: within the core radius ($R_{\rm C}$), between the core and the half-mass radius ($R_{\rm C-HM}$), and outside the latter ($R_{\rm oHM}$). Moreover, the binary fraction, $f$, measured in any cluster region was separately estimated in three ranges of the binary mass ratio: q$>0.5$ [f(q$>0.5$)], q$>0.6$ [f(q$>0.6$)], and q$>0.7$ [f(q$>0.7$)]. Formally, the total binary fraction f(tot), which takes the expected (not measured) fraction of binaries with q$\leq0.5$, is given, too. However, this parameter was calculated by the authors from  f(q$>0.5$) by accepting one of the existing versions of the mass-ratio distribution of binary stars. So, the former is simply equal to the latter multiplied by a numerical coefficient. Therefore, f(q$>0.5$) may be used as f(tot) whenever the absolute value of f(tot) is unimportant. Unfortunately, there is a lack of binary measurements within the $R_{\rm C}$ region in a notable number of the sample GCs, namely in 16 clusters being primarily more massive and concentrated globulars, such as NGC 104 (47 Tuc), NGC 1851, NGC 5904 (M 5), NGC 6093 (M 80) NGC 6388, NGC 7078 (M 15) among others. Fortunately, 11 of these GCs have measurements in the $R_{\rm C-HM}$ region where the populations of binaries still remains large. \citet{milonetal2012a} find that the fraction of binaries, on average, decreases by a factor of $\sim 2$ from the center to about two core radii. For this reason, in order to obtain both a uniform and large sample size as possible, we used a larger set of GCs where binary star fractions were measured just in the $R_{\rm C-HM}$ region. Overall, 45 GCs forming this data set turned out to be in common between the lists of \citet{Milone_etal2017} and \citet{milonetal2012a}. The GCs of this sample fall in a large range of the key GC parameters, such as metallicity, mass, and stellar density. 

\section{RESULTS}\label{results}

Due to high stellar densities, the central regions of GCs are characterized by unique conditions, which favor frequent close encounters between stars and important effects caused by such encounters. In particular, in relation with the subject of the present paper, these events dynamically affect hard and soft binaries in such a way that may favor them to finally merge\slash collide or disrupt, respectively (see more detail in next Subsection~\ref{bifratio_g1} and Section~\ref{discussion}). We analyze the above-described data on GCs and their multiple stellar populations to further examine one of the key aspects of the suggested collisional\slash merging formation of G2 stars presently observed in GCs, i.e. with mass $M_{MS} < 0.9$ M$_{\sun}$, of primordial low-mass MS (G1) stars. Namely, the present paper looks into the role of (initial) binaries and respective implications.
  
\subsection{The fraction of G1 stars as a function of binary star fraction} \label{binfrac_g1}

\begin{figure*}
\includegraphics[angle=-90,width=1\textwidth]{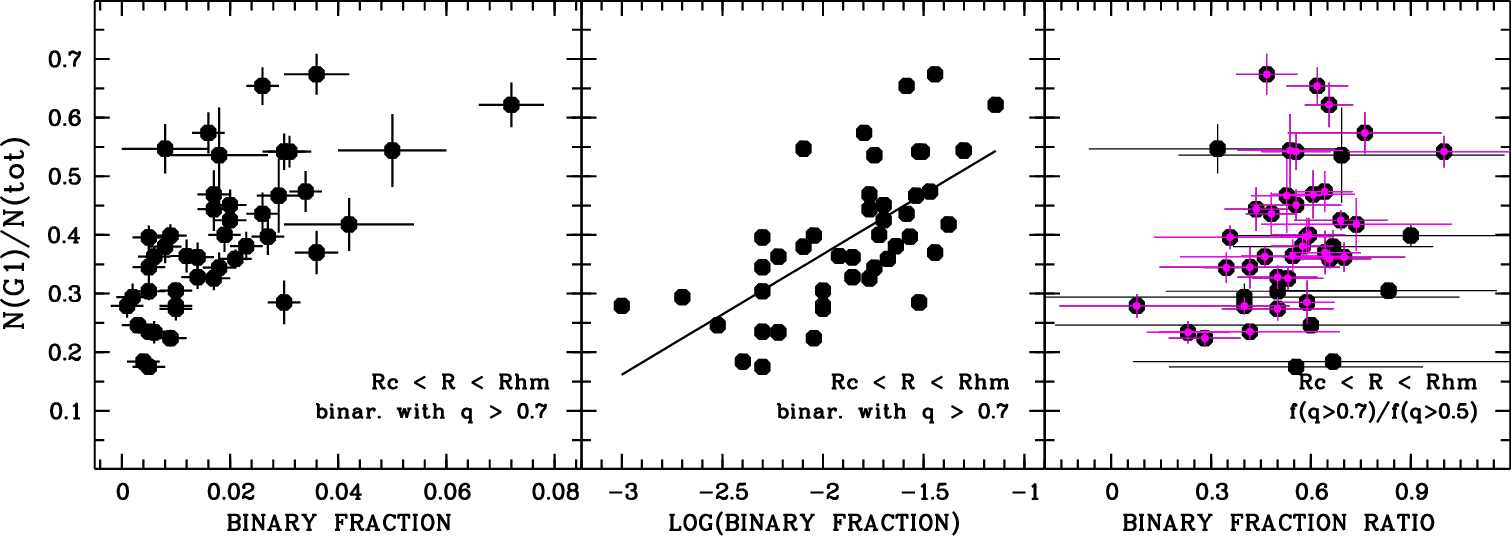}
\caption{For a sample of 45 Galactic GCs: The dependencies of the fraction of G1 stars ($N_{\rm G1}/N_{\rm tot}$) on (a, left panel) the fraction of binary stars with the mass ratio q$>0.7$, f(q$>0.7$), isolated and quantified by \citet{milonetal2012a} in the cluster area between the core and the half-mass radius ($R_{\rm C} < R < R_{\rm HM}$); (b, middle panel) the logarithm of f(q$>0.7$); (c, right panel) the ratio between the fractions of binary stars with the mass ratio q$>0.7$ and q$>0.5$, f(q$>0.7$)$/$f(q$>0.5$). The continuous line in middle panel is a linear fit to the plot. Black points in the right panel are 10 GCs (22\% of the sample) with the biggest errors of the ratio f(q$>0.7$)$/$f(q$>0.5$) (see more details in the text).}
\label{fig:fig1}
\end{figure*}

The crucial role of binary-mediated collision\slash merging of G1 stars in the formation of (the bulk of) G2 stars in GCs implies that it is a binary consuming process. Hence, one may expect a relationship between the observed binary fraction and the population of G1 stars. Figure~\ref{fig:fig1} displays the dependence of the parameter ($N_{\rm G1}/N_{\rm tot}$), the fraction of G1 stars, on f(q$>0.7$), the fraction of binaries with the mass ratio q$>0.7$ in GCs of the sample. In the middle panel of Figure~\ref{fig:fig1}, the parameter ($N_{\rm G1}/N_{\rm tot}$) is plotted as a function of the logarithm of f(q$>0.7$). The continuous line in the middle panel is a linear fit to the data. The plot in the right panel of Figure~\ref{fig:fig1} shows the parameter ($N_{\rm G1}/N_{\rm tot}$) as a function of the ratio between the fractions of binaries with different mass ratios, f(q$>0.7$) and f(q$>0.5$). This dependence is discussed in detail in Subsection~\ref{bifratio_g1} along with a summary of closely related relevant results obtained and accumulated to date on binary stars.

We estimated the level of statistical significance of the dependencies shown in Figure~\ref{fig:fig1}. The results of the Spearman's rank correlation test done to assess the correlation between the data sets plotted in Figure~\ref{fig:fig1} are listed in Table~\ref{tab:table_1}. As can be seen, the correlation is statistically significant at very high confidence level for the same dependence shown in two forms (left and middle panels). We made separate estimates for both forms of the dependence in order to take care of a formality. The significance of the trend in the right panel was estimated for two its versions, the total and reduced. The reduced version is coded by magenta color in Figure~\ref{fig:fig1}. It is obtained by reducing from the total set of data (45 units) 10 GCs (22\% of the sample) with the largest errors of the ratio f(q$>0.7$)$/$f(q$>0.5$). These points make the trend more diffuse and somewhat less significant according to the test. The reduced version of the trend is surely significant at a high confidence level of 99.91\% as compared to somewhat lower confidence level of 97.07\% for the total data set.

These dependencies provide another piece of evidence supporting our scenario. First, the fraction of MS binaries is proportional to the fraction of G1 stars in GCs or, in other words, this means that the smaller fraction of binary stars in GCs, the larger, on average, the fraction of G2 stars. A systematically decreasing fraction of binaries in GCs with systematically increasing fraction of G2 stars may be explained as due to the formation of G2 stars from binaries. However, it is known \citep{milonetal2012a} that GCs with larger fraction of binaries are also systematically more massive. So, the discussed relation may alternatively be interpreted as being due to more efficient disruption of binaries in more massive GCs. The trend presented in the right panel of Figure~\ref{fig:fig1} does not support the latter interpretation. The trend shows that the smaller the fraction of G1 stars is, the more deficient binaries of higher mass ratio (q$>0.7$) are. These binaries of higher mass ratio are, on average, harder than their smaller mass-ratio counterparts due to higher binding energy at a given primary mass. In the next Subsection, we justify this in detail.

\begin{table}
	\caption{Spearman's rank correlation test.}
	\label{tab:table_1}
\begin{tabular}{cllc}
		\hline
Dependence & $N_{\rm GC}$ & $\rho^a$ & P(\%)$^b$ \\
		\hline
($N_{\rm G1}/N_{\rm tot}$) vs f(q$>0.7$) & 45 & 0.674 & $>99.9999$\\
($N_{\rm G1}/N_{\rm tot}$) vs log f(q$>0.7$) & 45 & 0.674 & $>99.9999$\\
($N_{\rm G1}/N_{\rm tot}$) vs f(q$>0.7$)$/$f(q$>0.5$) & 45 & 0.325 & 97.07\\
($N_{\rm G1}/N_{\rm tot}$) vs f(q$>0.7$)$/$f(q$>0.5$) & 35 & 0.537 & 99.91\\
		\hline
\end{tabular}\\

\footnotetext{a}{$^a$ Spearman's $\rho$ coefficient of correlation between ($N_{\rm G1}/N_{\rm tot}$) and other quantities characterizing binary stars in GCs.}\\
\footnotetext{b}{$^b$ The probability that two data sets are associated.}

\end{table}

\subsection{The fraction of G1 stars as a function of the relative fraction of higher mass-ratio binaries} \label{bifratio_g1}

This subsection deals with the subject of the most suitable and probable kind of binaries that could primarily be involved in the (trans)formation of (G1 into) G2 stars. The dependence of ($N_{\rm G1}/N_{\rm tot}$) on the binary fraction ratio (right panel, Figure~\ref{fig:fig1}) implies that it is their fraction with q$>0.7$ of the mass-ratio distribution. Indeed, this dependence unambiguously shows that the relative fraction of the higher mass-ratio binaries decreases with decreasing fraction of G1 RGB stars in GCs. Virtually the same binaries are expected to be the hardest in GCs at a given primarily mass. Strictly speaking, the data on the fractions of MS binary stars of different mass ratio in GCs used in our analysis do not allow one to distinguish between soft and hard binaries. So, one cannot directly refer to the kind of these binaries, as soft or hard. Then, do we have any ground to relate the obtained dependencies with our approach? In other words, may we interpret the formally decreasing amount of binaries with decreasing fraction of G1 stars in GCs as probably due to the transformation of the binaries into G2 stars rather than to the disruption of a larger number of binaries in GCs with smaller ($N_{\rm G1}/N_{\rm tot}$)? Of course, the issue is really complex. Nevertheless the trend in the right panel of Figure~\ref{fig:fig1} supports such an interpretation, given a body of both observational and theoretical present-day data about binaries and multiple systems. In order to explain this trend, we briefly summarize key actual data on binary stars and their characteristics appropriate to the case, particularly on hard (close) binaries, their probable mass-ratio distribution and their fate in GCs.

According to the present-day understanding of the issue, the dynamical evolution of binaries (triples) in GCs develops in two opposite ways, namely: dynamically hard and soft binaries become even harder and softer, respectively \citep[e.g.,][and references therein]{2022Speretal,RoznerPerets2022}. Thus, dynamically soft binaries can hardly survive for a long time in the high-density environment of GCs. In particular, both tidal disruption by the gravitational potential of GCs and dynamical evaporation tend to soften and finally disrupt these binaries \citep{RoznerPerets2022}. Therefore, a fraction of the initially formed binaries, primarily soft ones, are expected to be essentially disrupted, while on the other hand, hard binaries increase their binding energy. There is a well known complexity of the effects on hard binaries in the condition of high-density environment due to their interactions with other GC constituents, such as single stars, other binary stars, and black holes. A part of their increasing binding energy is lost to kinetic energy of escaping single stars. Also, one cannot exclude that hard binaries achieving a threshold binding energy can eject themselves before merging. Nevertheless, despite these effects, the most probable development and the fate of the dynamical evolution of hard binaries in GCs is expected to be merger or collision \citep{GoodmanHut1993}. Supporting evidence of a similar effect with a lower efficiency is observed among field stars. \citet{Tokovininetal2006} established that close spectroscopic binaries with and without tertiary companion have significantly different period distributions whereas their mass-ratio distributions are similar. So, the dynamical effect of such tertiary companions leads to shorter periods of respective binaries. It is now evident that virtually all short-period binary stars in the field have a tertiary companion. In the more general context of the observed relation between close binaries and solar-type hierarchies, \citet{Tokovinin2023} finds that the inner subsystems in these hierarchies show strong excess of periods shorter than 30 days. Interestingly enough, the eccentricities of the same subsystems tend to be zero, and a larger fraction of twins belongs to these short-period binaries, too. 

In addition to the dynamical effects making hard binaries become harder in the dense environment of GCs, with their probable fate being a merger, there may be other potential effects favoring the merging of binaries. For instance, \citet{RoznerPerets2022} argue that the properties of binaries in the early GCs could be affected by the gas produced by the evolved G1 stars. The presence of gas can lead to an efficient merger of binaries, on the one hand, and to the formation of additional population of binaries, on the other.

It is now widely accepted that the mass-ratio distributions of MS binaries in both the field and clusters are in obvious disagreement with random pairing of stars taking their proportions from the IMF into account. Moreover, a body of observational evidence converge towards a particular feature in the mass-ratio distribution of solar-type binaries in the form of a prominent peak (excess) of twin binaries, i.e. with mass-ratio close to unity. We primarily refer just to low-mass, so-called solar-type stars, with the primary mass $0.5-1.5$ M$_{\sun}$, which are relevant to this study. \citet{Fisheretal2005} studied a sample of field spectroscopic MS binaries with the primary mass greater than $1.1$ M$_{\sun}$ and found an evident peak in the mass-ratio distribution near q=1. Using a sample of 454 solar-like (F6–K3) stars from the Solar vicinity, \citet{Raghavanetal2010} revealed that the mass-ratio distribution shows "a preference for like-mass pairs", which are more frequent in relatively close pairs. \citet{TokovininMoe2020} compared their model with observational data on binary stars, namely with a large sample of solar-type binaries obtained by \citet{Tokovinin2014a} within 67 pc from the Sun. They noted that the model reproduces well the large excess of twin binaries (with q$>0.95$), the fraction of which is larger among close binaries \citep{Tokovinin2014b}. In the old open cluster M67, \citet{AlbrowUlusele2022} estimated the fraction of photometric binaries with q$>0.8$ to be $\sim$14\% (and $\sim$26\% with q$>0.5$). From their overview of a variety of observational studies on binary stars including low-mass ones, \citet{PinsonnStanek2006} argued that there is clear evidence of a typically large fraction of twin binaries (or at least with q$>0.8$) revealed in these works (up to 35\% of such binaries). 

By proceeding from the observational evidence that close binaries typically have mass ratios close to unity, the so-called twin binaries, \citet{Adamsetal2020} theoretically explained this as being due to the fact that this corresponds to the lowest energy state of such binary systems.

\section{DISCUSSION AND CONCLUSION} \label{discussion}

The initial star formation and the dynamical processes acting in the dense stellar environment of GCs objectively define both the fraction and mass-ratio distribution of the hard binaries which are suggested to be the direct progenitors of G2 stars. It cannot be excluded that the mass-ratio distribution of binaries in this particular environment of GCs can deviate from its version in the Galactic field in the sense that the former distribution is likely more weighted toward higher mass ratio so that the majority of hard binaries fall between $0.8<$ q $<1.0$. 

\subsection{The formation of G2 stars and quantitative estimate of the fraction of G1 stars at a given mass} \label{specified_approach}

We qualitatively formulate our approach in a more specific form defined by the transformation of high mass-ratio G1 binaries into G2 stars of approximately two times higher mass than the primary one. This means that lower-mass G1 stars forming such twin or almost twin binaries are transformed into a more massive G2 stars, which replace less numerous (in absolute value, but at the same or comparable fraction) G1 binaries at this higher mass, since these latter stars are in turn transformed into even more massive G2 stars. This way, a successive shifting of more numerous (in absolute value) lower-mass stars, transformed into G2 stars, towards two times more massive (and therefore many times less numerous at that mass) G1 stars is a mechanism that can naturally and uniformly operate and lead to similar outcome in different GCs.  

We now attempt to obtain more quantitative estimates of how the binary star fraction can relate to the fraction of G1 stars. We use the following arguments. Suppose that the fraction of hard binaries of G1 stars, $f_{\rm B,M}$, corresponding to the number $N_{\rm B,M}$, with both the primarily mass around $M$ and high mass ratio (virtually twins), are transformed into the same number of G2 stars with mass around 2$M$. Similarly, due to the same process, $b$ times larger number of hard binaries with the primary mass around $0.5 M$ are transformed into single G2 stars, $N_{\rm S,M}({\rm G2})$, with mass around $M$, while the number of G1 stars left around the same mass $M$ is $N_{\rm S,M}({\rm G1})$. In other words, the transformed $N_{\rm B,M}$ G1 binaries were substituted with $b$ times larger number of G2 stars, $N_{\rm S,M}({\rm G2})$. Then, $N_{\rm S,M}({\rm G2})=b N_{\rm B,M}$. Note that the fraction of hard binaries $f_{\rm B,0.5M}$ is not necessary equal to $f_{\rm B,M}$ in general case, since the fraction of hard MS binaries in a GC may vary along the MS. One can easily qualitatively express the binary fraction $f_{\rm B,M}$ the following way:

\begin{equation}
f_{\rm B,M}=\frac{N_{\rm B,M}}{N_{\rm S,M}({\rm G1})+N_{\rm B,M}},
\label{eq1}
\end{equation}

Our goal is to approximately estimate a typical realistic value of the parameter $g_{\rm 1}$, the fraction of G1 MS stars around mass $M$, which was defined in \citet{Kravtsovetal2022} as the analog of the fraction of G1 RGB stars [i.e., $N_{\rm 1}({\rm G1})/N_{\rm tot}=N_{\rm 1}({\rm G1})/(N_{\rm 1}({\rm G1})+N_{\rm 2}({\rm G2}))$]. Taking the above-mentioned expressions into account, $g_{\rm 1}$ is expressed as:

\begin{equation}
g_{\rm 1}=\frac{N_{\rm S,M}({\rm G1})}{N_{\rm S,M}({\rm G1})+N_{\rm S,M}({\rm G2})}=\frac{N_{\rm S,M}({\rm G1})}{N_{\rm S,M}({\rm G1})+b N_{\rm B,M}},
\label{eq2}
\end{equation}

From here and using Equation~\ref{eq1} we easily come to the following expression:

\begin{equation}
g_{\rm 1}=\frac{1-f_{\rm B,M}}{1+f_{\rm B,M}(b-1)}
\label{eq3}
\end{equation}

For estimating typical value of $g_{1}$, the value of $b$ may be around $b\approx5$, i.e. equal to the ratio of the number of single stars, with the given ratio of their mass, defined by the canonical Kroupa IMF \citep{Kroupa2001,Kroupa2002}\footnote{This value of $b$ can be different if the shape of the IMF is calibrated according to the IMF slope versus stellar surface density of the cluster found by \citet{Dib2023}.}. Namely, we accept the ratio of their mass is equal to $\approx 2$ and the slope of the IMF is $\alpha = 2.3$ in the range of low-mass stars, at $M_{\rm MS} < 1.0$ M$_{\sun}$. Since the fraction of hard, high-mass ratio (virtually twin) binaries that have finally merged in GCs is an obvious variable, we consider two distinct values for it, namely $f_{\rm B,M}=0.10$ and $f_{\rm B,M}=0.30$. They presumably bracket the plausible range for this variable in GCs. Each of the two values is accepted to be constant in the considered stellar mass range, although, in the general case, the fraction of hard (close) binaries may vary with stellar mass as we noted earlier. Then, by applying Equation~\ref{eq3}, we obtain the values of $g_{\rm 1}$ equal to 0.64 and 0.32, respectively. This is a reasonable range in comparison with the fractions of the G1 RGB stars, ($N_{\rm G1}/N_{\rm tot}$) that are estimated for most GCs in the Milky Way. However, massive GCs have lower values of ($N_{\rm G1}/N_{\rm tot}$), typically between 0.20 and 0.25. Such values of $g_{\rm1}$ can be obtained for some useful and interesting cases. In particular, it is worth noting the aforementioned case of $f_{\rm B,M}$ varying along MS, for example, as an increasing function of decreasing stellar mass. This tendency seems to be real at least for the field stars at solar metallicity. For instance, \citet{TokovininMoe2020} note an interesting detail, namely that "the excess twin fraction is substantially reduced for close early-B binaries and quickly diminishes with increasing separation" in contrast with the same characteristic estimated by the same authors for solar-type stars. In such a case, suppose that the value of $b$ increases, say, up to $b=7$ between the same masses, $0.5M$ and $M$, within $M_{\rm MS} < 1.0$ M$_{\sun}$. Then the value of $g_{\rm 1}$ decreases up to $g_{\rm 1}=0.25$ for $f_{\rm B,M}=0.30$. Note, however, that a detailed analysis of this issue is outside the scope of the present paper.

\subsection{The compatibility of a merger-induced formation of G2 stars with their surface abundance of light elements} \label{CNOcyc}

The mechanism described above for the formation of G2 stars directly suggests that the abundance of light elements (primarily CNO-Na) distinguishing G2 stars from their G1 counterparts, were synthesized in the stellar interiors and transported to the stellar surface. However, this point is very disputable and it suffers from high grade of uncertainty. The formation of a merger-induced G2 MS star as a result of the merger of two G1 MS stars has yet to be investigated in detail. Its particularly challenging step is the formation of one whole hydrogen-burning nucleus from the two merging ones. The merging of G1 stars is accompanied with the transformation of the system angular momentum and with an initially strongly perturbed non-equilibrium state of the merger product. The coalescence of the hydrogen-burning nuclei of G1 stars into one whole nucleus should temporally boost the central temperature and pressure in the forming nucleus as compared to the same parameters in the nucleus of a G1 star of the same mass. As of now, there is unfortunately significant uncertainty about this process and a lack of reliable knowledge about the temporal variations of the central temperature, its maximum possible value achievable at the initial stage of the formation of the merger product. This would have a strong impact on the yields of relevant nuclear reactions. Whether the difference of the CNO-Na surface abundances between G1 and G2 stars can be explained in the framework of our scenario due to the processes generated by low-mass mergers in the interiors of their products is a key question that should be clarified by dedicated numerical simulations. The results obtained so far on the modeling and simulation of the processes related to the mergers of low-mass stars do not exclude or even partially support the ant-correlated variation of the surface abundances of at least some elements. Such a variation of N and C may take place, increasing with time, but emerging at best $\sim2$ Gyr after the event occurred, depending on the model \citep{Sillsetal2005}.

According to the models of \citet{Prantzosetal2007}, the core H-burning temperature in stars increases towards the end of the MS evolution and it may reach up to $\sim27\times10^6$K even in low-mass stars with  $M_{\rm MS} = 0.9$ M$_{\sun}$ (at [Fe/H]=-1.6). The authors note that this temperature is yet somewhat lower than what is typically required for the NeNa-chain to operate (it becomes sufficient at the RGB, particularly near its tip, where the temperature can exceed $\sim60\times10^6$K). In their Figure 4 (left panel), \citet{Prantzosetal2017} indicate the temperature range as a function of the produced He amount ($\Delta Y$, which is used as a proxy for time) where each of the relevant chemical elements is either overproduced (including Na) or depleted with respect to its initial value by the factors corresponding to the extreme abundances observed in the GC NGC 2808 where [Na/Na$_0$]$ > 0.7$. The low temperature limit necessary for synthesizing the required amount of Na exactly for this particular case is around $40\times10^6$K at $\Delta Y < 0.02$. The lower the value of [Na/Na$_0$] is, the lower is the temperature limit.

We only briefly note that in the framework of the merger\slash collision formation of G2 stars, a qualitative interpretation of the disappearing Mg-Al anti-correlation in less massive or most metal-rich GCs can be considered in terms of the expected impact of decreasing velocity dispersion and increasing viscosity of the stellar matter, respectively. Indeed, the kinetic energy of colliding\slash merging H-burning nuclei apparently depends on the two factors at a given stellar mass.  

In this context, it is important to point out to another interesting advantage of the merger-induced scenario for the formation of G2 stars as compared with their canonical formation from H-processed material ejected by polluters (and diluted with pristine gas with composition of G1 stars). The former mechanism requires an order of magnitude less amount of the enriching material to form the same amount (i.e., the same total mass) of G2 stars with a given chemical composition. The main reason is that this difference is defined by the much lower efficiency of star formation of the canonical mechanism. It can hardly be more than 20\% as compared with the almost 100\% efficiency ($\sim$90\%) of the merger-induced mechanism. In other words, this mechanism requires to synthesize, say, at least 5 times less amount of the required chemical elements. This directly implies an apparently lower required rate of the involved nuclear reactions corresponding to a lower temperature limit. Whether the central temperature could temporally achieve or even exceed the minimum required level in a forming low-mass merger\slash collision products of $M_{\rm MS} = 0.8-0.9$ M$_{\sun}$ (the bulk of the available spectroscopic data on the O, Na, Mg, Al abundances in GCs were obtained in GCs for RGB stars) is one of the key question that have yet to be answered.

In contrast to the high degree of uncertainty of the possibility for the necessary amount of material to be processed at least through the CNO cycle and the NeNa chain in low-mass MS collision products, the availability of a mechanism capable of transporting such a material to the external parts of these products (MS G2 stars) looks very likely. Indeed, the effects of rotation in the interiors of collision products are theoretically known more reliably, with lower degree of uncertainty. Collision products are expected to be fast rotators at least in the initial period after their formation. \citet{Lombardietal1996} used smoothed particle hydrodynamics simulations of stellar collisions between MS stars in GCs and noted that any off-axis collisions result in rapidly, differentially rotating merger remnants. Later, \citet{Sillsetal2005} confirmed that these off-axis collision products have substantial angular momentum from their initial configuration (even assuming initially non-rotating stars). Moreover, \citet{sills2015a} noted that very high initial rotation rates of the collision products must be reduced by an unknown process to allow these products to collapse to the MS. So, rotational mixing is the very likely mechanism that would transport the yields of the assumed CNO cycle to the stellar surface. This process and its impact on the variation of the surface elemental abundance of collision products is considered, in particular, in the models of low-mass collision products by \citet{Sillsetal2005}. Their models show, for example, essential variation of the surface abundance of N (in excess) and C (in depletion). Add also that one of the main effects of tides in close binaries is that they tend to spin-up the companions and this tidal spin-up establishes rotational mixing in the stellar interiors of the companions before their merging \citep{2022Speretal}.

\subsection{A possible effect of merging on the surface Li abundance} \label{lithium}

The proposed mechanism of the formation of G2 stars not only avoids the problematic mass budget problem common to other alternatives, but also it apparently does not need an external primordial gas for dilution. {We have to note, however, that according to \citet{Prantzosetal2007}, in the case of the dilution of Li-free gas processed through the CNO cycle at high temperature in the polluters (which are massive or intermediate-mass stars in the scenarios relying on the canonical formation of G2 stars, though), the situation is different, i.e. "the unprocessed material has to be of pristine nature (i.e. pure ISM) rather than from the stellar envelope of the polluter itself".} 

On the other hand, rotational mixing, as the main mechanism responsible for the variation of the surface elemental abundances of merger products, should not lead to an essential depletion in the surface lithium abundance of the merger products (G2 stars). \citet{Strassmeieretal2012} find that binaries exhibit, on average, approximately 0.25 dex less surface Li than single stars of comparable temperature, as expected if the depletion is due to the rotation. In turn, \citet{AguileraGomezetal2023} find no evident impact of binarity on Li abundance among field RGB stars of their sample. Even more so \citet{Sayeedetal2023} find a larger fractions of fast rotators among Li-rich stars. The issue of Li abundance in low-mass stars, enriched by numerous observational results for the last years, is contradictory and even surprising. Recent studies of low mass-stars sometimes show paradoxical results on Li abundance. \citet{Sayeedetal2023} find that Li-rich stars at the base of the RGB have somewhat higher mean s-process abundances and confirm a prevalence of Li-rich stars on the red clump, i.e. among core helium-burning red giants. Surprising enough that this enrichment with Li, not predicted by a standard evolution theory, is somehow related with the helium flash, since it is observed among stars with mass $M < 2$ M$_{\sun}$ \citep{Kumaretal2020}. According to \citet{Lietal2018}, Li-rich stars are found on the MS. They argue that this indicates the presence of an unknown efficient process that affects the surface Li abundances before the RGB stage. In addition, Li abundance is a function of various parameters. For instance, \citet{Martosetal2023} reveal robust anti-correlations between Li abundance and both metallicity and age among solar twins.

Perhaps the most relevant result to mergers between non-compact stars has been recently obtained by \citet{Kaminsketal2023} on the presence of lithium in three Galactic red nova remnants, which are probably produced by mergers of MS stars. Super-solar lithium abundances they find in these objects "may suggest that at least some merger products activate mixing mechanisms capable of producing lithium". V1309 Sco is among these red novae. Its outburst was detected and studied photometrically. It has been established to be caused by the merged stars of a contact binary with the mass of the main companion $\sim 1$ M$_{\sun}$ \citep{tylendaetal2011} or perhaps somewhat higher, $\sim 1.5$ M$_{\sun}$. This is now the least massive binary system known to give rise to a red novae \citep{howittetal2020}. This object and studies on its nature were discussed in more detail in \citet{Kravtsovetal2022}.

\subsection{Concluding remarks} \label{remarks}

We obtained new results and complemented them with both a body of observational evidence and theoretical results on binary stars in different environments including GCs. This allowed us to tentatively identify the probable progenitors of G2 stars in the framework of our approach that suggests the binary-mediated merging\slash collision formation of G2 stars in GCs. These progenitors are high mass-ratio (virtually twin) hard binaries in GCs. They form in any environment, but only in the condition of the high density environment in GCs, and possibly in some other objects, such as ultra-compact dwarf galaxies, the bulge of the Galaxy, etc., these "seeds" are converted into G2 stars.

The consumption of G1 binaries for building G2 stars means that G1 and G2 sub-populations should be deficient and virtually free of primordial binaries, respectively. This is in agreement with observations that show that the fractions of binaries among G1 sub-populations in GCs are normally an order of magnitude higher than among G2 ones \citep[see][and references therein]{Grattonetal19}.

Primordial (G1) MS stars composing low-mass binaries also undergo an essential transformation due to mass transfer between the components, particularly the accretor. The transferred gas not only increases the accretor's mass but it can also modify the surface elemental abundances under certain conditions. For example, by using the MESA code, \citet{Weietal2020} carried out model studies of surface chemical anomalies of low-mass MS stars due to mass transfer. Their results show that in some low-mass binaries, the accretors can exhibit elemental abundances similar to those of G2 stars, i.e. C and O are in depletion whereas Na and N are in excess (at constant Mg and Al, however), while still remaining on the MS. Note, however, that the donor is considered to be at least as massive as $1$ M$_{\sun}$ and the final result depends on a few other key parameters including the mass ratio that is typically adopted to be q$=0.5$. When the accretor evolves and is on the RGB, these abundance anomalies can become weaker or disappear entirely. Therefore, given these model results, one can note two useful points. First, the mass-transfer binaries that recently were at the stage of blue stragglers and now are (their accretors) at the RGB, are expected to originally be of systematically lower mass ratio as compared to their counterparts, hard binaries with q$ >0.8$ that merged a long ago (G2 stars) and actually are also on the RGB. Second, how many of the actually observed RGB stars in GCs are the evolved blue stragglers? \citet{Weietal2021} estimated that their fraction is about 10\%. Our estimate \citep{Kravtsov_Calderon2021} made for the GC NGC 3201 taking the observed number of blue stragglers in this GC cluster into account is of the same order of magnitude, possibly slightly smaller. However, strictly speaking, this fraction depends on the specific frequency of blue stragglers recently ($\sim 1$ Gyr ago) populated GCs. Normally, the specific frequency of the presently observed blue stragglers is higher in GCs with higher fractions of binary stars \citep[see][and references therein]{milonetal2012a,Chatteretal2013}.

By considering the particular example of NGC 3201, \citet{Kravtsov_Calderon2021} paid attention that the present-day RGB stars in GCs are not uniform in terms of their mass, since not all of them are the progenies of the primordial MS stars. One (minor) fraction are the relatively recently evolved blue stragglers of any nature (with mass typically $M_{\rm BS} > 1.1$ M$_{\sun}$) and another fraction of RGB stars are expected to be the progenies of merger\slash collision products of the primordial MS stars (latter, in \citet{Kravtsovetal2022} and in the present paper we argued that the bulk of G2 stars can be such products). By relying on the models of \citet{Sills_etal2009}, \citet{Kravtsov_Calderon2021} argued also that such products should achieve the RGB and evolve on it together with the progenies of the primordial stars of (typically) lower mass. In this context, it is worth noting recent results on estimating the masses of stars belonging to the advanced stages of evolution, mainly the RGB, by using asteroseismic observations of the GCs M4 \citep{Howelletal2022} and M80 \citep{Howelletal2023}. Several outliers with masses exceeding $M_{\rm RGB} > 1$ M$_{\sun}$ were detected as candidates evolved blue stragglers. Moreover, an increased dispersion of the derived masses of RGB stars, excluding the outliers, at and below the RGB bump allows to assume a real variation of mass of the order of $\sim 0.05$ M$_{\sun}$.

The present paper aims at developing further our recently proposed approach \citep{Kravtsovetal2022} on the merger\slash collision nature of G2\footnote{In the present paper, we use the terms multiplicity, first and second generations (G1 and G2) to follow the definitions adopted in the widely accepted paradigm of multiple stellar populations in GCs, which unambiguously implies the canonical mode of the formation of G2 stars from the gas produced by G1, (very) massive and intermediate-mass stars. Perhaps, G1 and G1$_{tr}$ (transformed) instead of G1 and G2 (or P1 and P2), respectively, would be more appropriate for our conceptual framework.} stars in GCs. The new results reported in the present paper are added with a variety of important both observational evidence and theoretical results obtained so far in various works on this and closely related subjects. Thereby, we further argue a new way of understanding the origin of G2 stars.

\section{Acknowledgments}

The authors thank the anonymous referee for both the expressed interest to the results obtained in the present paper and for useful comments that improved its final version. 

\section*{Data Availability}

The data underlying this article are available in tables from the papers referred to here, at https://doi.org/10.1051/0004-6361/201016384 and https://doi.org/10.1093/mnras/stw2531.

\bibliographystyle{mnras}
\bibliography{paper}

\label{lastpage}

\end{document}